\documentclass[a4paper]{article} 
\usepackage{graphicx}
\usepackage{amsmath,amssymb}

\begin{document}
\thispagestyle{empty} \parskip=12pt \raggedbottom
 
\def\mytoday#1{{ } \ifcase\month \or January\or February\or March\or
  April\or May\or June\or July\or August\or September\or October\or
  November\or December\fi
  \space \number\year}
\noindent
\vspace*{1cm}
\begin{center}
  {\LARGE New results on cut-off effects in spectroscopy with the fixed point 
action} 
\vskip7mm 
 Peter Hasenfratz$^a$, K.~Jimmy Juge$^b$ and Ferenc Niedermayer$^{a,c}$ \\
BGR (Bern-Graz-Regensburg) Collaboration 
\vskip6mm
$^a$
  Institute for Theoretical Physics, 
  University of Bern \\
  Sidlerstrasse 5, CH-3012 Bern, Switzerland
\vskip1mm

$^b$
  School of Mathematics,
  Trinity College \\
  Dublin 2, Ireland
\vskip1mm 
$^c$
On leave of absence from E\"otv\"os University, HAS Research Group,
Budapest, Hungary
\vskip6mm

  \nopagebreak[4]
 
\begin{abstract}
Our study on the cut-off effects in quenched light hadron spectroscopy 
and pion scattering length with the fixed point action is extended by 
results obtained at a lattice spacing a=0.102 fm in a box of size L=1.8 fm. 
The cut-off effects are small, but clearly seen as the resolution is 
increased from a=0.153 fm to a=0.102 fm. 
In the quark mass region where the errors are small and under control, 
our results on the APE plot lie close to the extrapolated numbers 
of the CP-PACS Collaboration.
\end{abstract}
 
\end{center}
\eject

\section{Introduction and summary}
In a recent paper\cite{BGR}, the BGR-Collaboration presented results on light 
hadron (down to $m_\pi/m_{\rm N} \approx0.25$) quenched spectroscopy with 
two different actions: the 
parametrized  fixed point (FP) action\cite{fpg,fpd} and the CI action which 
is a combination of 
the L\"uscher-Weisz gauge action\cite{LuWe} with a chirally improved Dirac 
operator\cite{CI}. In this paper we extend the scaling analysis in 
ref.\cite{BGR} on the FP action by adding new results on an $18^3 \times 36$ 
lattice with lattice unit $a(r_0)=0.102$\,fm. This extension is necessary,
since our scaling test in \cite{BGR} referred to a small $L=1.2$\,fm spatial
size lattice with low statistics, while our large 
$L=2.4$\,fm lattice results with good statistics could be compared with other 
simulations only. The new data have been used already in a scaling test on the
$I=2, \pi\pi$ scattering length\cite{JJ}. No cut-off effects were seen there,
but it is difficult to measure the scattering length precisely.  
We need reliable information on the cut-off effects 
in the quenched approximation since these results help to choose the
parameters of future full QCD simulations the BGR Collaboration is moving 
towards. In addition, this information might be helpful when comparing the
results from different actions for consistency. Indeed, there exist
unclarified inconsistencies between different works in light hadron
spectroscopy. On the APE plot, for example, the CP-PACS\cite{CP-PACS}
continuum extrapolated numbers lie significantly below the data obtained with
most of the different improved actions.
For a recent comparison of numerical data obtained with different actions
see ref.~\cite{compare}.

The parametrized FP Dirac operator lives on the hypercube, has 81 offsets, 
a large number of paths, uses all the elements of the Clifford 
algebra\cite{fpd} and satisfies the Ginsparg-Wilson relation up to
parametrization errors. The setup of the present simulation and the analysis
follows closely those applied in ref.\cite{BGR}. We generated 180 gauge 
configurations separated by 500 alternating Metropolis and pseudo 
over-relaxation sweeps at gauge coupling $\beta=3.4$. At this coupling 
the lattice unit is $a(r_0)=0.102$\,fm as determined from the Sommer 
parameter $r_0=0.49$\,fm\cite{Sommer}. 
As it is well known, it is difficult to give an estimate on the systematic
errors of $a(r_0)$ if the lattice is not fine. 
At this resolution, the systematic error should be on the percent 
level, while the statistical error is smaller than $1\%$\cite{fpg}. The
spatial extension of the lattice is $L\approx 1.8$\,fm. In our earlier results
obtained at $L=1.8$\,fm and $L=2.4$\,fm at resolution $a=0.153$\,fm we saw
no finite size effects beyond the statistical errors. Our scaling analysis
might be influenced by small finite size effects, however, and these physical
effects might mix with the topological finite size artifacts as we discuss 
below.

Before summarizing our results, let us make two general remarks concerning
scaling analyses in light hadron spectroscopy. 
Although the most interesting part of the spectrum for physics is 
where the quark masses go to their small physical values, the heavier 
quarks bring more information for scaling studies: 
the hadrons are more compact, they are more 
difficult to resolve and the cut-off effects are expected to be larger.
In addition, the statistical and systematic errors are smaller, so the
analysis is more conclusive. 

\begin{figure}[htb]
\begin{center}
\includegraphics[width=0.8\textwidth]{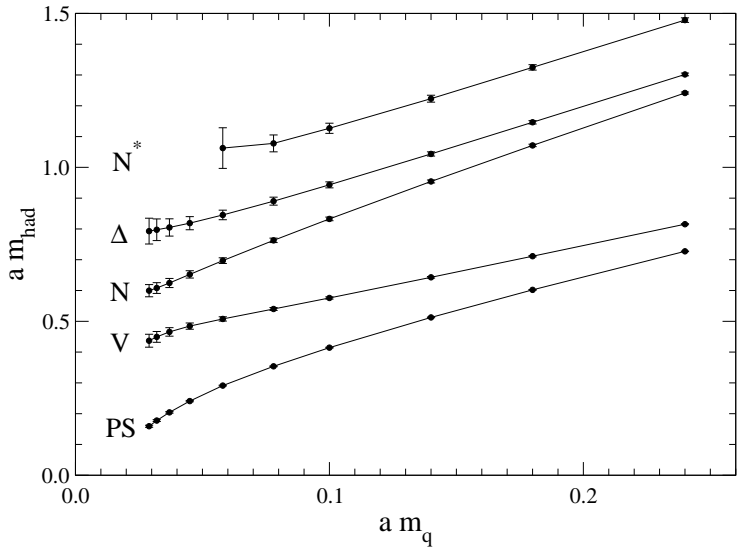}
\end{center}
\vspace{-5mm}
\caption{{}\label{fig:rawdata} The hadron masses as the function
of the quark mass in lattice units at $a=0.102$\,fm in a box of size
$L=1.8$\,fm.} 
\end{figure}

The second remark refers to a special artifact of {\it quenched} light hadron
spectroscopy: the topological finite size artifacts. The discrete zero modes of
the Dirac operator at $m_q=0$ are not suppressed in the quenched approximation
and, in a finite volume at sufficiently small quark masses, they will corrupt
the hadron propagators\cite{topartifact}. In ref.\cite{BGR} we
eliminated/reduced the topological finite size 
artifacts by using special hadron correlators in the pseudoscalar (PS), 
nucleon (N) and delta ($\Delta$) channels. Unfortunately, in the vector
channel we did not find a natural solution for this problem and the artifacts
turned out to be larger at $a(r_0)=0.102$\,fm than at $a(r_0)=0.153$\,fm
in the same physical volume $L=1.8$\,fm. The effect is seen in 
the vector meson mass at the lightest quark masses in Fig.~\ref{fig:rawdata},
where the dimensionless hadron masses are shown as the function of the
dimensionless quark mass. This feature (namely, the bending down of the vector
meson curve at small quark masses) is very similar to the artifacts 
we saw earlier appearing in smaller volumes, and then  disappearing in our
largest box with $L=2.4$\,fm\cite{BGR}. These quenching 
artifacts should be separated from the cut-off effects in a scaling test.
This is relevant in particular in the vector channel, since the vector meson
mass is used traditionally (unfortunately) as a unit for other hadron
masses. 
The problem occurs at sufficiently light quark masses only, which gives 
an additional argument to avoiding this region in scaling studies.

We give here a brief summary of our conclusions.
Considering mass ratios
(like in the APE plot, or measuring the masses in units of a well determined
fixed hadron mass) we see small cut-off effects in the baryon channel when
comparing the results at $a=0.153$ and 0.102\,fm resolutions. 
The effect is seen for relative heavy  quarks beyond the statistical errors. 
In the
$N^*$ channel this is a several standard deviation effect and has a size of
up to $\approx 2\%$ in the ratio $m_{N^*}/m_V$, where $m_V$ is the vector
meson mass. In the nucleon and delta channels the cut-off effects are smaller,
but for heavy quarks they are also visible beyond the statistical errors.
This small shift brings the FP results on the APE plot closer to 
and essentially consistent with the CP-PACS
continuum extrapolated results with Wilson fermions\cite{CP-PACS}. The
improved staggered (asqtad\cite{asqtad}) data at $a=0.13$\,fm\cite{MILC} 
with heavier quarks lie a few
standard deviations higher and are close to the CP-PACS results at
$a=0.05$\,fm. 
In addition, unlike at $a=0.153$\,fm, we see a discrepancy in 
the overall scale obtained in the gauge sector from the Sommer parameter 
$r_0$ and the scale from the spectrum of light hadrons. 
The hadron sector prefers a scale of $a=0.106(2)$\,fm as opposed 
to the scale of $a(r_0)=0.102$\,fm. In this case, however, it is 
difficult to estimate the systematic errors.  

\section{Scaling analysis}
Concerning the details of the analysis we refer to ref.\cite{BGR}. In order
to avoid very busy figures we shall use a few selected results only from
other works in this paper. 
One can get a more complete picture by combining the figures here 
with those in ref.\cite{BGR}.

\begin{figure}[htb]
\begin{center}
\includegraphics[width=0.8\textwidth]{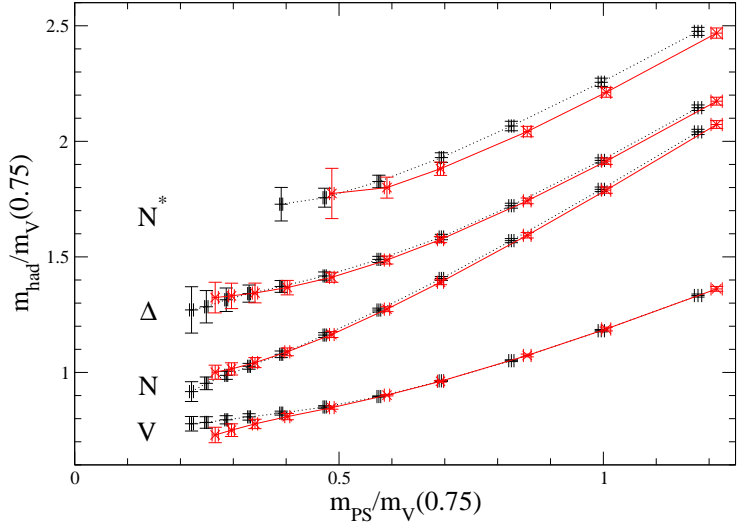}
\end{center}
\vspace{-5mm}
\caption{{}\label{fig:overview} The hadron masses in the vector meson,
  nucleon, $\Delta$ and $N^*$ channels as the function of the pseudoscalar
  mass. All the masses are measured in $m_{\rm V}(0.75)$ units. 
  Here  $m_{\rm V}(0.75)$ is the vector meson mass at a quark mass value, 
  where $x= m_{\rm PS}/m_{\rm V}=0.75$. The points connected by dotted and
  continuous lines  refer to $a(r_0)=0.153$\,fm and 0.102\,fm, respectively. }
\end{figure}

The measured new data are collected in Tables 1 and 2.
Fig.~\ref{fig:overview} gives a quick overview on the cut-off effects in the
different channels. In this figure the hadron masses are measured in 
$m_{\rm V}(0.75)$ units, where $m_{\rm V}(0.75)$ is the vector meson mass at
$x=m_{\rm PS}/m_{\rm V}=0.75$. This figure indicates already that the change
due to cut-off effects is small in $a \in (0.153-0.102)$\,fm.   

\begin{table}[htb]
\begin{center}
\renewcommand{\arraystretch}{1.2} 
\begin{tabular}{c|c|c|c|c|c} 
\hline
$am_q$ & $am_\mathrm{PS}$ & $am_\mathrm{V}$ & $am_\mathrm{Oct}$(N) &
  $am_\mathrm{Dec}$(N) & $am_\mathrm{Oct}(\mathrm{N}^*$) \\
\hline
 0.029 &  0.1592(33) & 0.437(21) & 0.599(20) & 0.793(42) &           \\
 0.032 &  0.1776(29) & 0.449(18) & 0.608(17) & 0.797(35) &           \\
 0.037 &  0.2042(25) & 0.466(14) & 0.624(15) & 0.805(28) &           \\
 0.045 &  0.2408(21) & 0.484(10) & 0.652(12) & 0.819(21) &           \\
 0.058 &  0.2913(14) &  0.508(7) & 0.697(9)  & 0.845(15) & 1.063(66) \\
 0.078 &  0.3537(13) &  0.540(5) & 0.763(7)  & 0.890(13) & 1.078(27) \\
 0.100 &  0.4144(12) &  0.576(4) & 0.833(6)  & 0.943(10) & 1.127(17) \\
 0.140 &  0.5127(11) &  0.643(3) & 0.954(5)  & 1.044(7)  & 1.223(11) \\
 0.180 &  0.6021(10) &  0.711(2) & 1.071(4)  & 1.146(5)  & 1.325(9)  \\
 0.240 &  0.7272(9)  &  0.815(2) & 1.241(3)  & 1.302(4)  & 1.478(7)  \\
\hline
\end{tabular}
\end{center}
\caption{{} The hadron masses for different quark masses in lattice units. 
 \label{table:1}}
\end{table}

\begin{table}[htb]
\begin{center}
\renewcommand{\arraystretch}{1.2} 
\begin{tabular}{c|c|c|c|c|c} 
\hline
$x$ & $\frac{m_\mathrm{N}(x)}{m_\mathrm{V}(x)}$   & 
      $\frac{m_\Delta(x)}{m_\mathrm{V}(x)}$       & 
      $\frac{m_\mathrm{V}(x)}{m_\mathrm{V}(x_0)}$ & 
      $\frac{m_\mathrm{N}(x)}{m_\mathrm{V}(x_0)}$ & 
      $\frac{m_\Delta(x)}{m_\mathrm{V}(x_0)}$     \\
\hline
0.40  & 1.351(64)  & 1.769(97) & 0.753(37) & 1.018(27)  & 1.333(54) \\
0.45  & 1.340(41)  & 1.719(65) & 0.784(26) & 1.051(22)  & 1.348(41) \\
0.50  & 1.347(27)  & 1.689(44) & 0.810(18) & 1.092(18)  & 1.368(32) \\
0.55  & 1.364(19)  & 1.672(31) & 0.834(12) & 1.138(15)  & 1.395(25) \\
0.60  & 1.385(14)  & 1.659(23) & 0.864(8)  & 1.196(12)  & 1.433(20) \\
0.65  & 1.410(10)  & 1.649(19) & 0.898(5)  & 1.266(10)  & 1.481(18) \\
0.70  & 1.436(8)   & 1.641(14) & 0.940(2)  & 1.350(7)   & 1.543(13) \\
0.75  & 1.461(6)   & 1.633(10) & 1.000(0)  & 1.461(6)   & 1.632(10) \\
0.80  & 1.486(5)   & 1.623(7)  & 1.079(2)  & 1.602(6)   & 1.750(10) \\
0.85  & 1.507(4)   & 1.610(5)  & 1.199(5)  & 1.808(8)   & 1.931(10) \\
\hline
\end{tabular}
\end{center}
\caption{{} Hadron masses in $m_{\rm V}(x)$, or $m_V(x_0)$ units, where
$x=m_{\rm PS}/m_{\rm V}$ and $x_0=0.75$.  
 \label{table:2}}
\end{table}

\begin{figure}[htb]
\begin{center}
\includegraphics[width=0.8\textwidth]{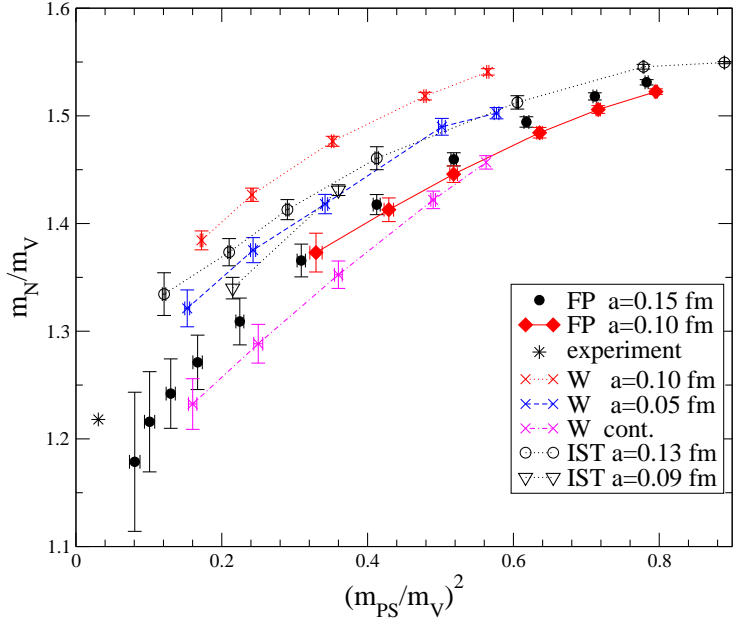}
\end{center}
\vspace{-5mm}
\caption{{}\label{fig:APE} APE plot, where the FP numbers are compared to the
  unimproved Wilson fermion data and their continuum extrapolation. The
  improved staggered (asqtad) results are shown also.   }
\end{figure}

\begin{figure}[htb]
\begin{center}
\includegraphics[angle=270,width=0.99\textwidth]{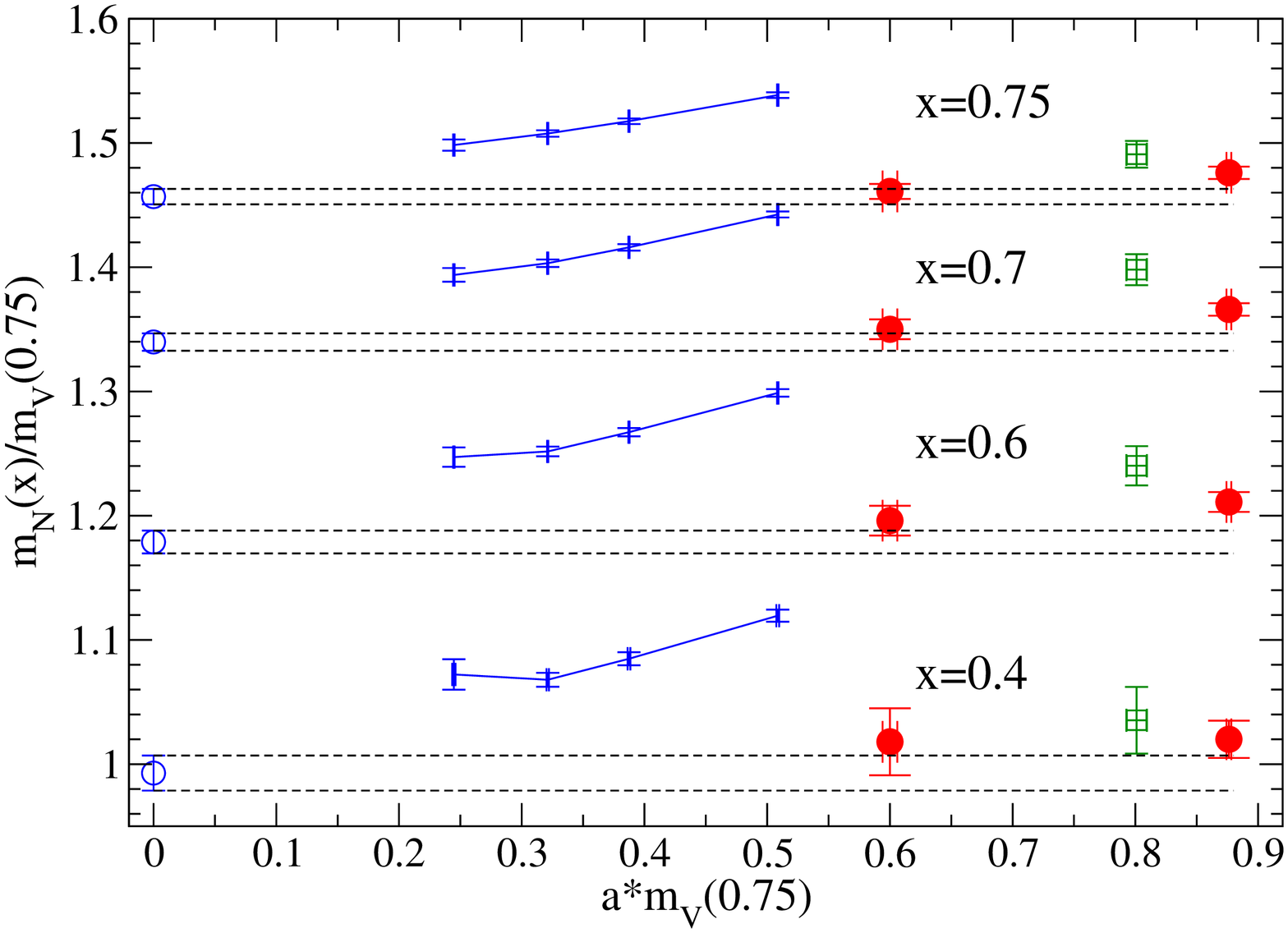}
\end{center}
\vspace{-5mm}
\caption{{}\label{fig:FP-CI-CP} The cut-off dependence of the nucleon mass in 
 $m_{\rm V}(0.75)$ units for different $x= m_{\rm PS}/m_{\rm V}$
values. The CP-PACS Wilson data (4 points for $a \in (0.10,0.05)$\,fm and their
continuum extrapolation (circles at $a=0$), the FP numbers at $a=0.153$ and 
 $a=0.102$\,fm and the CI results at $a=0.148$\,fm (squares) are 
compared in this figure.}
\end{figure}

Fig.~\ref{fig:APE} shows $m_N/m_V$ as the function of 
$(m_{\rm PS}/m_{\rm V})^2$ (APE plot). For the new $a(r_0)=0.102$\,fm FP
data the 4 lightest points are not shown due to the problems in the vector
meson channel discussed above. The remaining points
lie slightly below of, but within the errors are consistent with our earlier 
$a(r_0)=0.153$\,fm results. If anything, the new data moved slightly
towards the CP-PACS continuum extrapolated Wilson data\cite{CP-PACS}. This
is illustrated also by Fig.~\ref{fig:FP-CI-CP}, where 
$m_{\rm N}(x)/m_{\rm V}(0.75)$ is shown as the function of 
$a m_{\rm V}(0.75)$ for different fixed $x=m_{\rm PS}/m_{\rm V}$
values. Here, $m_{\rm V}(0.75)$ is the vector meson mass at $x=0.75$ which is
used as a convenient mass unit to measure the nucleon mass $m_{\rm N}(x)$.
In this figure the CP-PACS data obtained in the 
$a=0.10-0.05$\,fm range and their continuum extrapolations are compared to the
FP data at $ a(r_0)=0.153$\,fm and $ a(r_0)=0.102$\,fm. 
Also included in this plot are the ci results at $a(r_0)= 0.148$\,fm.
The Wilson action has large cut-off effects and the continuum extrapolation is
a non-trivial task, even when the finest lattice spacing is 0.05\,fm.
Nonetheless, our results seem to support the continuum extrapolated
CP-PACS numbers, in particular at $x=0.75$, where the errors are small. The
$x=0.4$ new FP data on this figure should be taken cautiously due to the fact
that $m_N(x=0.4)$ is indirectly influenced by the topological artifact 
problem in the vector channel. In fig.~\ref{fig:APE}
the improved staggered (asqtad) data lie several standard deviations higher and
are close to the CP-PACS Wilson action results at $a=0.05$\,fm. 

\begin{figure}[htb]
\begin{center}
\includegraphics[width=0.8\textwidth]{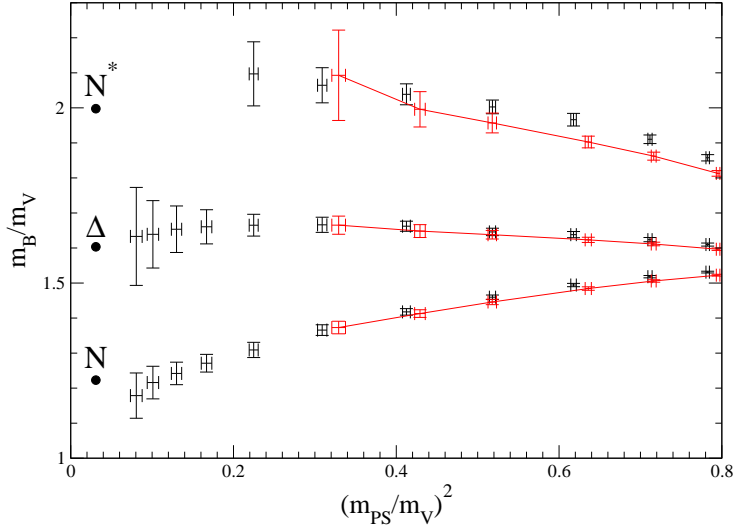}
\end{center}
\vspace{-5mm}
\caption{{}\label{fig:NDNstar} APE plot with the FP results at $a=0.102$\,fm
  (connected by a continuous line) and  $a=0.153$\,fm in the $N, \Delta$ and
  $N^*$ channels. }
\end{figure}

\begin{figure}[htb]
\begin{center}
\includegraphics[width=0.8\textwidth]{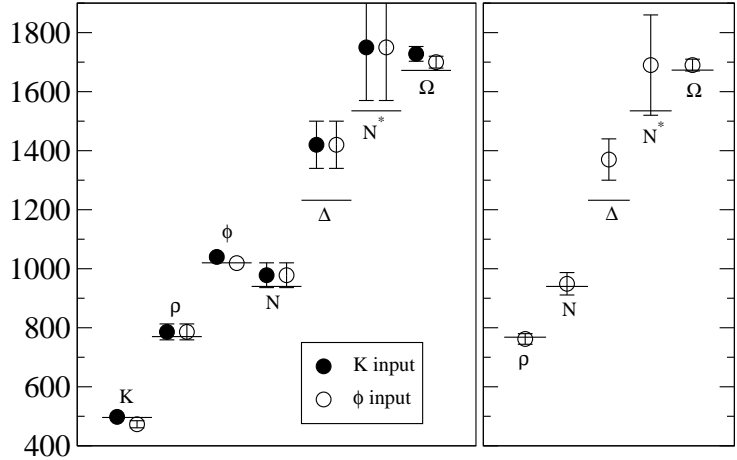}
\end{center}
\vspace{-5mm}
\caption{{}\label{fig:spectrum} Hadron masses in MeV units. The lattice unit
  and the quark masses $a,m_{ud}$ and $m_s$ are fixed by $r_0, \pi, K$, or
  $\Phi$ (on the left) and by the $\pi, K$ and $\Phi$ mesons (on the right),
  respectively. }
\end{figure}

Fig.~\ref{fig:NDNstar} is an APE plot again where the FP data at 
$a(r_0)=0.153$ and 0.102\,fm are compared for the 
nucleon ($N$), delta ($\Delta$) and the negative parity baryon $N^*$. The
nucleon, as we discussed above, and the $\Delta$ show some shift downwards as
the lattice resolution is increased, 
but this effect is hardly statistically significant.
This shift is, however, clearly seen for the $N^*$. We conclude
that the FP action has an observable resolution error at $a(r_0)=0.153$\,fm
for this heavy, negative parity state.

Fig.~\ref{fig:spectrum} shows the hadron spectrum fixing the lattice unit and
quark masses $a,m_{ud}$ and $m_s$ by $r_0, \pi, K$, or $\Phi$ (on the left)
and by the $\pi, K$ and $\Phi$ mesons (on the
right), respectively. Extrapolating (interpolating) the measured
hadron masses to their physical points we followed the procedure in
ref.\cite{BGR} using functional forms dictated by quenched chiral perturbation
theory\cite{qch}. 
On the r.h.s. the stable $N$ and narrow $\Omega$ are at the right place, 
whereas the broad $\Delta$ and $N^*$ are pushed upwards. 
The $\rho$ is also at the correct place, but only by accident since 
the vector meson mass is pushed down by the topological finite size 
artifacts at light quark masses as we discussed before.
The fitted value of the lattice unit $a=0.1055(14)$\,fm (the error is
statistical only) deviates from the scale found in the gauge sector
$a(r_0)=0.102$\,fm (error on the percent level). 
On the l.h.s. the scale is taken from $r_0$ in the gauge sector. This spectrum
is somewhat distorted which comes from the difference between the preferred
scale in the hadron spectrum and in the gauge sector from $r_0$.
 
\begin{figure}[htb]
\begin{center}
\includegraphics[width=0.8\textwidth]{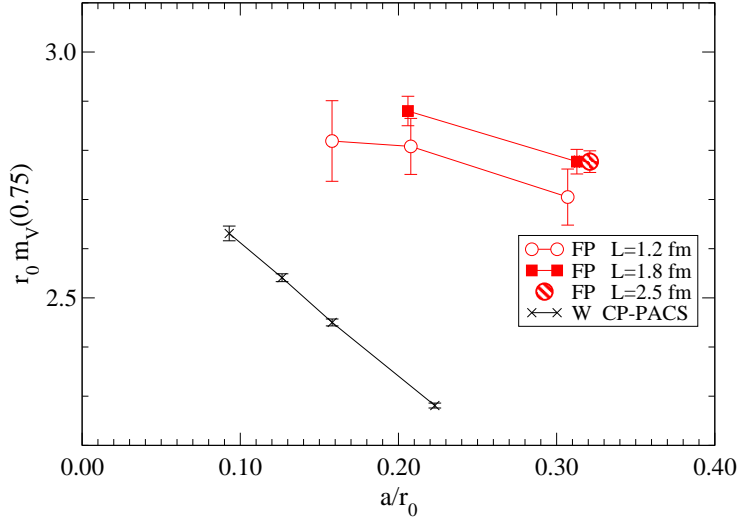}
\end{center}
\vspace{-5mm}
\caption{{}\label{fig:r0mV} Scaling behavior of $m_V(0.75)$ in $1/r_0$ unit.}
\end{figure}

In Fig.~\ref{fig:r0mV} the product $r_0 m_V(0.75)$ is shown as the function of 
$a/r_0$ with $r_0$ from the gauge sector. The new data confirm the trend
indicated by the $L=1.2$\,fm data in ref.\cite{BGR}: $r_0 m_V(0.75)$ is
increasing as we go towards the continuum limit. This effect is seen in the
new data beyond the statistical errors. The value at $a(r_0)=0.102$\,fm 
seems to be consistent with the continuum extrapolated CP-PACS number. 

\section{Pion Scattering Length}

The pion scattering length (in the isospin 2 channel) was studied as an 
application other than standard hadron spectroscopy. We use the finite 
volume method of L\"uscher~\cite{Luscher} to relate the energy shift 
of two pions in a finite volume to the infinite volume scattering length,
\[
\delta E=-\frac{4\pi a_0}{m_\mathrm{PS} L^3}
\left\{1+c_1\frac{a_0}{L}+c_2\frac{a_0^2}{L^2}\right\}+\dots
\]
where $c_1 = -2.837297$ and $c_2 = 6.375183$. 
The fitting procedure is described in ref.~\cite{JJ}.

Cutoff effects were studied in the small volume $(1.2\,\rm{fm})^3$ using all
three different lattice spacings ($0.08, 0.10, 0.15\,\rm{fm}$) and in an
intermediate volume of $(1.8\,\rm{fm})^3$ with $\rm{a}\sim0.10$ and 
$0.15$\,fm. 
A wide range of quark masses were studied, with the smallest quark mass
resulting in $m_\mathrm{PS}/m_\mathrm{V}\sim0.35$. In this quenched study, we
keep those masses with $m_\mathrm{PS} L\gtrsim4$. 

\begin{figure}[htb]
\begin{center}
\includegraphics[width=0.8\textwidth,height=0.6\textwidth]{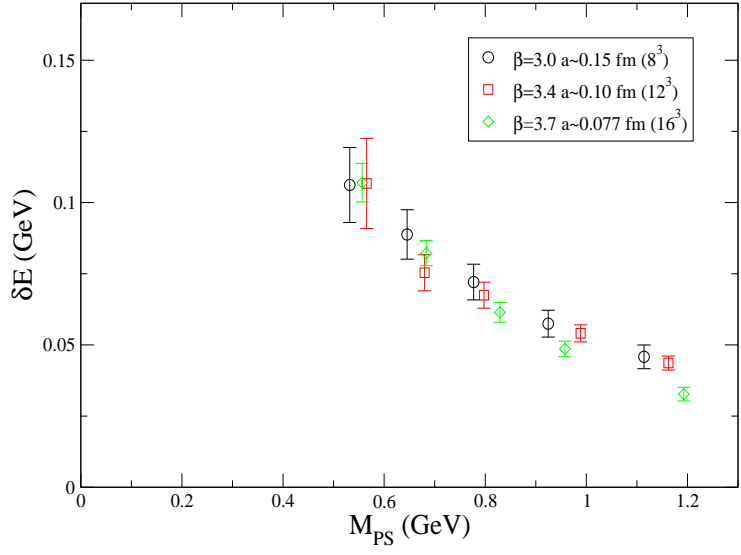}
\end{center}
\vspace{-5mm}
\caption{{}\label{fig:small_vol_pipi} The energy shift of the two pion state 
in the $(1.2\,{\rm fm})^3$ box for the three different lattice spacings.}
\end{figure}

In the small volume, the translation of the energy shift to the scattering
length may involve large corrections and so it is preferable to simply compare
the energy shifts rather than the scattering lengths. In
Fig.~\ref{fig:small_vol_pipi}, we show the energy shift of the two pion state
in the $(1.2\,{\rm fm})^3$ box for all three different lattice spacings. The
energy shift below a mass of 1 GeV shows scaling within the errors from the
three different lattice spacings. 

\begin{figure}[htb]
\begin{center}
\includegraphics[width=0.8\textwidth,height=0.6\textwidth]{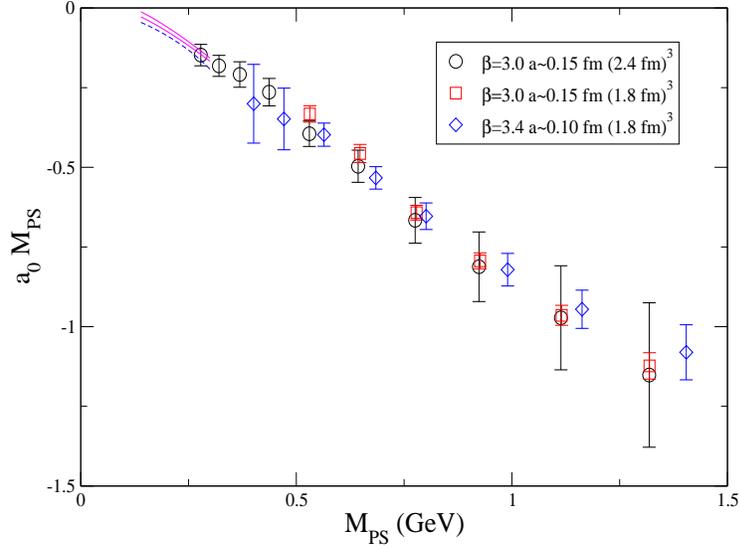}
\end{center}
\vspace{-5mm}
\caption{{}\label{fig:large_vol_pipi} The dimensionless scattering length in
  the $(1.8\,{\rm fm})^3$ box for two different lattice spacings, ${\rm
    a}=0.10\,{\rm fm}$ and ${\rm a}=0.15\,{\rm fm}$. Chiral perturbation
  theory curves (\cite{qXpt,BernX}) are shown for $m_\mathrm{PS}<0.3\ {\rm
    GeV}$.}
\end{figure}

For the intermediate volume, ($1.8\,{\rm fm})^3$, we calculate the scattering
length from L\"uscher's formula up to ${\mathcal O}(1/L^5)$ where the
pseudoscalar mass satisfies $m_\mathrm{PS} L\gtrsim4$. The dimensionless
quantity $a_0m_\mathrm{PS}$ is shown in Fig.~\ref{fig:large_vol_pipi}. Scaling
is observed in this quantity below a pseudoscalar mass of $\sim1.2\ {\rm
  GeV}$. For masses greater than $1.2\ {\rm GeV}$, it is possible that some
scaling violations appear, but the differences are still consistent within the
statistical errors. We include in Fig.~\ref{fig:large_vol_pipi} the results
from the largest volume to show that the truncation at $1/L^5$ is a negligible
effect and that the extracted scattering length is indeed a physical quantity
(within the quenched approximation). The scattering lengths are listed in
Tables~\ref{table:pipi16}-\ref{table:pipi34}.

\begin{figure}[htb]
\begin{center}
\includegraphics[width=0.8\textwidth,height=0.6\textwidth]{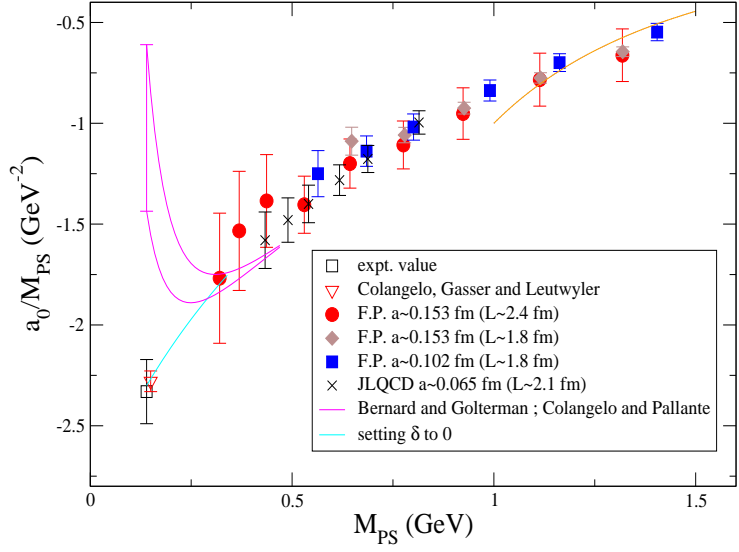}
\end{center}
\vspace{-5mm}
\caption{{}\label{fig:summary} Summary of all the intermediate and large
  volume results together with the the Wilson action results (\cite{JLQCD})
  for the ratio $a_0/m_\mathrm{PS}$. NNLO chiral perturbation theory results
  (\cite{BernX}) and NLO quenched chiral p.t. curves (\cite{qXpt}) are shown
  with the measured values for $F_\pi$. The behaviour expected from the
  scattering of hard spheres are shown at large $m_\mathrm{PS}$.}
\end{figure}

We compare our results from large and intermediate volumes to other large
scale simulations using the Wilson action. The comparison requires some care
as the volumes used are not always the same. However, we have seen that the
truncation is under control at least if the volume is greater than 
$(1.8\,{\rm fm})^3$. The physical quantity that remains finite in 
the chiral limit, $a_0/m_\mathrm{PS}$, is plotted against 
$m_\mathrm{PS}$ in physical units in
Fig.~\ref{fig:summary}. We note that, in addition to the agreement of the
$a=0.15$\,fm and $a=0.10$\,fm results, all of our scattering lengths are
statistically consistent with the finest lattice results from the JLQCD
collaboration \cite{JLQCD} who use the standard Wilson action at
$\beta=6.3$. We conclude that the cutoff effects of the FP action on the pion
scattering length are small at nearly twice the lattice spacings required for
unimproved actions. This gain is important for such quantities where a chiral
action in a large volume is needed.

\begin{table}[htb]
\begin{center}
\renewcommand{\arraystretch}{1.2} 
\begin{tabular}{c|c|c|c|c} 
\hline
 $am_q$ & $a\delta E$ & $a_0/a$ & 
   $a_0/(a^2m_\mathrm{PS})$ & $a_0m_\mathrm{PS}$\\
\hline
0.33  &  0.0041(10)  & -1.12(22)  & -1.10(22) & -1.15(23)\\
0.25  &  0.0049(10)  & -1.12(19)  & -1.30(22) & -0.97(16)\\
0.18  &  0.00597(97) & -1.13(15)  & -1.58(21) & -0.81(11)\\
0.13  &  0.00691(89) & -1.11(12)  & -1.84(20) & -0.666(72)\\
0.09  &  0.00733(90) & -0.99(10)  & -1.99(20) & -0.497(50)\\
0.06  &  0.0085(10)  & -0.958(97) & -2.33(23) & -0.395(40)\\
0.04  &  0.0081(15)  & -0.78(13)  & -2.30(38) & -0.264(43)\\
0.028 &  0.0089(19)  & -0.73(14)  & -2.54(49) & -0.209(40)\\
0.021 &  0.0103(21)  & -0.73(13)  & -2.93(54) & -0.182(33)\\
\hline
\end{tabular}
\end{center}
\caption{{} The scattering lengths for the $\beta=3.0,\ 16^3\times32$ lattice. 
 \label{table:pipi16}}
\end{table}

\begin{table}[htb]
\begin{center}
\renewcommand{\arraystretch}{1.2} 
\begin{tabular}{c|c|c|c|c} 
\hline
 $am_q$ & $a\delta E$ & $a_0/a$ & $a_0/(a^2m_\mathrm{PS})$ & 
 $a_0m_\mathrm{PS}$\\
\hline
0.33  &  0.01021(49) &  -1.096(40) & -1.070(40) & -1.123(41)\\
0.25  &  0.01232(52) &  -1.113(37) & -1.285(43) & -0.964(31)\\
0.18  &  0.01466(59) &  -1.103(35) & -1.533(48) & -0.793(25)\\
0.13  &  0.01660(76) &  -1.062(38) & -1.755(64) & -0.642(23)\\
0.09  &  0.0164(13)  &  -0.908(57) & -1.81(11)  & -0.457(29)\\
\hline
\end{tabular}
\end{center}
\caption{{} The scattering lengths for the $\beta=3.0,\ 12^3\times24$ lattice. 
 \label{table:pipi30}}
\end{table}

\begin{table}[htb]
\begin{center}
\renewcommand{\arraystretch}{1.2} 
\begin{tabular}{c|c|c|c|c} 
\hline
 $am_q$ & $a\delta E$ & $a_0/a$ & $a_0/(a^2m_\mathrm{PS})$ & 
 $a_0m_\mathrm{PS}$\\
\hline
 0.240 &  0.00562(56) &  -1.49(12)  & -2.04(16) &  -1.080(86) \\
 0.180 &  0.00728(58) &  -1.57(10)  & -2.61(16) &  -0.945(60) \\
 0.140 &  0.00877(69) &  -1.60(10)  & -3.12(19) &  -0.821(51) \\
 0.100 &  0.01062(85) &  -1.576(99) & -3.80(24) &  -0.653(42) \\
 0.078 &  0.0117(10)  &  -1.50(10)  & -4.25(28) &  -0.533(35) \\
 0.058 &  0.0126(14)  &  -1.36(12)  & -4.66(43) &  -0.398(36) \\
 0.045 &  0.0160(55)  &  -1.43(40)  & -5.9(1.7) &  -0.348(97) \\
 0.037 &  0.0191(96)  &  -1.45(60)  & -7.0(2.8) &  -0.30(12)  \\
\hline
\end{tabular}
\end{center}
\caption{{} The scattering lengths for the $\beta=3.4,\ 18^3\times36$ lattice. 
 \label{table:pipi34}}
\end{table}

\section{SUMMARY}

\begin{figure}[htb]
\begin{center}
\includegraphics[width=0.8\textwidth]{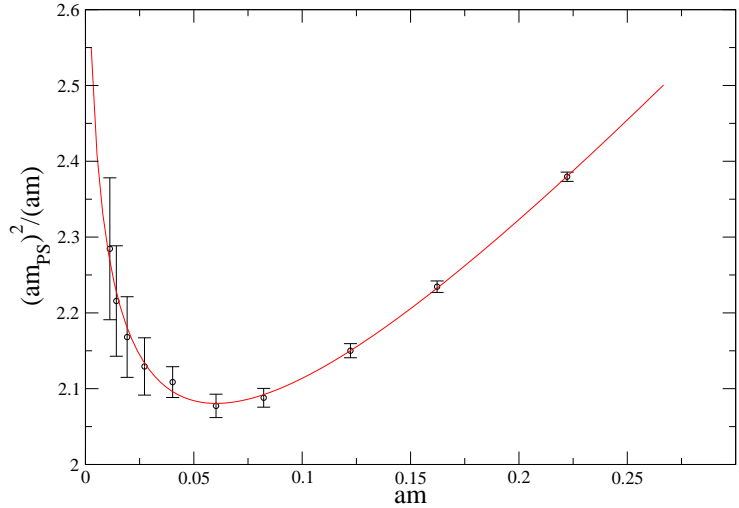}
\end{center}
\vspace{-5mm}
\caption{{}\label{fig:logdiv} The ratio $(am_\mathrm{PS})^2/(am)$ at fixed
  $a=0.102$\,fm as the function of the quark mass $am$, $m=m_q+m_{\rm
  res}$, where $m_{\rm res}$ is the residual additive renormalization due to
  the parametrization error of the FP Dirac operator. The continuous line is a
  fit of the form $\alpha+\beta\ln(am)+\gamma am$ suggested by quenched chiral
  perturbation theory.}
\end{figure}

In this work we discussed the cut-off effects in simulations with the
parametrized FP action. For this purpose going deep in the chiral limit is not
very important. It is good to see, nevertheless, that with this action one can
reproduce cleanly the divergence $\propto \ln(ma)$ in $(am_\mathrm{PS})^2/(am)$
predicted by quenched chiral perturbation theory\cite{qch} as shown in
Fig.~\ref{fig:logdiv}. Here $m=m_q+m_{\rm res}$, where $m_q$ is the quark mass
in the simulation and $m_{\rm res}$ is the residual additive mass
renormalization due to the parametrization error of the FP Dirac operator.  

{\bf Acknowledgements}
We thank Christine Davies for the correspondence on the interpretation of the
APE plot. We thank also the discussions with Anna Hasenfratz  and 
the members of the BGR Collaboration, and the support from 
the Swiss Center for Scientific Computing in Manno, 
where the numerical simulations were done. 
K.J.J. would like to thank Colin Morningstar for some of the fitting 
routines used in the analysis of the scattering length.
This work was supported by the Schweizerischer Nationalfonds.

\clearpage


\eject

\end{document}